\documentstyle[aps,preprint]{revtex}

\tightenlines
\begin{document}

\title{Avalanche exponents and
corrections to scaling for a stochastic sandpile}
			    
\author{Ronald Dickman$^{\dagger}$ and
J. M. M. Campelo
}
\address{
Departamento de F\'{\i}sica, ICEx,
Universidade Federal de Minas Gerais,
Caixa Postal 702,
30161-970 Belo Horizonte, Minas Gerais, Brazil\\
}
\date{\today}

\maketitle
\begin{abstract}
We study distributions of dissipative and nondissipative
avalanches in Manna's stochastic sandpile, in one
and two dimensions.  Our results lead to the following
conclusions: (1) avalanche distributions, in general, do not follow
simple power laws, but rather have the form
$P(s) \sim s^{-\tau_s} (\ln s)^{\gamma}
f(s/s_c)$, with $f$ a cutoff function; (2) the exponents
for sizes of dissipative avalanches in two dimensions
differ markedly from the corresponding values for the 
Bak-Tang-Wiesenfeld (BTW) model,
implying that the BTW and Manna models belong to distinct universality
classes; (3) dissipative avalanche distributions obey finite size 
scaling, unlike in the BTW model.
\vspace{1em}

\noindent$^\dagger${\small email: dickman@fisica.ufmg.br}

\end{abstract}


\date{\today}

%\section{Introduction}

Sandpile models are the prime examples of self-organized 
criticality (SOC) \cite{btw,dhar99}, 
a control mechanism that forces a system with an absorbing-state phase
transition to its critical point \cite{bjp,granada}, leading to
scale-invariance in the apparent absence of parameters \cite{ggrin}.  
Of central interest in the study of SOC are avalanche distributions,
which are expected to exhibit scale invariance.
It is generally assumed that the avalanche size and duration distributions
follow simple power laws in the infinite-size limit, and that
departures from such power laws reflect finite-size effects.
Such effects complicate the estimation of critical exponents,
since the estimates are senstive to the
choice of fitting interval.  

Recently, Drossel showed that in the Bak-Tang-Wiesenfeld (BTW) 
sandpile, distributions of {\it dissipative} avalanches 
(in which one or more particles leave the system),
follow clean power laws \cite{drossel}.  
Nondissipative avalanche distributions must also
follow power laws in the infinite-size limit \cite{drossel}, but are
subject to much stronger corrections to scaling.  The avalanche exponents
for the two cases are very different, and the proportion of
dissipative avalanches decreases $\sim L^{-1/2}$
with increasing system size $L$. 
Thus power-law
fits to the total avalanche distribution represent a superposition 
of two distinct scaling behaviors (with $L$-dependent weights), 
and would appear to have no fundamental significance.

In light of these findings, it is of interest to study dissipative and
nondissipative events separately in the stochastic sandpile as well.
Our principal results are that avalanche distributions for Manna's
sandpile are in general {\it not} pure power laws, but rather include
a logarithmic correction, and that the dissipative avalanche exponents
for the Manna model are quite different from those for the BTW model.
The latter serves to resolve the issue of distinct universality
classes for the two models, which has attracted considerable
attention \cite{chessa,biham}.

The version of the Manna sandpile \cite{manna}
studied here is defined on a hypercubic
lattice with open boundaries: a chain of $L$ sites in one dimension, 
a square lattice of $L \times L$ sites in $2d$. The
configuration is specified by the number of particles
$z_i $, at each site $i$; sites with $z_i \!\geq\! 2$ are 
{\it active}, and have a toppling rate of unity.  
When site $i$ topples, two particles move to randomly 
chosen nearest neighbors $j$ and $j'$ of $i$.  
($j$ and $j'$ need not be distinct.)
In 1d we report results for $L=500$, 1000, 2000, 5000,
and 10$\;$000 sites; in two dimensions the linear system sizes are
$L=160$, 320, 640, 1280, and 2560.  For the largest system sizes our
results are based on samples of about 10$^6$ avalanches, while for the
smallest systems about 10$^7$ avalanches are generated.
In 1d the simulations use sequential dynamics \cite{fes2};
in 2d studies with both parallel
and sequential updating were performed, yielding
identical results to within uncertainty.

We study the distributions $P_s(s)$ and $P_d(t)$
of avalanche sizes $s$ and durations $t$.
The data are binned to equal intervals of $\ln s$ and $\ln t$.
The fraction of dissipative avalanches decays $\sim L^{-1/2}$ in both
one and two dimensions.  (In 2d, for $L=2560$,
only 2\% of avalanches are dissipative.) 

The morphology of avalanche distributions in sandpiles generally
includes a plateau-like region for small $s$ or $t$, and a rapidly
decaying portion for large events; between these two limiting regimes
there is a power-law-like interval.  
The power-law interval 
is expected to grow with the system size, so that the probability 
distribution in the second and third regimes follows:
\begin{equation}
P_s(s) = s^{-\tau_s} f_s(s/s_c) ,
\label{usual}
\end{equation}
where $f_s$ is a scaling function that decays rapidly when its argument
is $\geq 1$, and the cutoff size $s_c \sim L^{D_s}$.  
For $s \ll s_c$, the scaling function takes a constant value $f_0$.
Analysis usually consists 
in selecting (in a plot of $P_s$ versus $s$ on log scales) a reasonably
linear portion, and performing a linear regression to the data to determine
$\tau_s$.
Fig. 1 illustrates the problematic nature of this procedure.  We have
plotted the size distribution of nondissipative avalanches in
two dimensions (with the parallel update scheme), along with a
polynomial fit to the data; the derivative of the latter, $-\tau$, is shown
in the inset.  Evidently we may have our choice of $\tau$ values
ranging from 1.25 to 1.41!

Any hope of extracting simple, power-law scaling from
a distribution of the kind shown in Fig. 1 (which is in fact typical
of both size and duration distributions for nondissipative avalanches,
regardless of system size or dimension), hinges on finding a 
suitable correction
to scaling term.  A natural choice, based on experience with critical
phenomena, would be to include a factor of the form $(1-as^{-\Delta})$ on the
r.h.s. of Eq. (\ref{usual}).  Attempts to fit such a form to the data
consistently yield values of $\Delta$ very near zero, suggesting instead
a logarithmic correction to scaling, so that Eq. (\ref{usual}) 
becomes
\begin{equation}
P_s(s) =  s^{-\tau_s} (\ln s)^\gamma f_s(s/s_c) .
\label{logcr}
\end{equation}
This expression may be further generalized by writing the
logarithmic term as $\ln (s/s_0)$; for reasons explained 
below we set $s_0 = 1$ in the present analysis.

We find that good fits to nondissipative avalanche distributions 
can only be obtained including the logarithmic correction.  Our
analysis consists in (1) making a preliminary estimate of the
fitting interval $[x_0,x_1]$ (here $x \equiv \ln s$); 
(2) adjusting parameters $\tau$ and $\gamma$
so as to minimize the variance of $f^* = s^\tau P_s(s)/(\ln s)^\gamma$
on the interval (ideally $f^*$ would be constant and the variance
zero); (3) checking for any systematic trend in $f^*$, and refining
the fitting interval accordingly.  In practice we use the largest
possible interval, excluding the small-$s$ plateau regime and the
large-$s$ cutoff.  For each kind of distribution, we use the 
same $x_0$ for each system size, while $x_1$ 
increases linearly with $\ln L$. 

Fig. 2 shows the result of this procedure, using $\tau = 1.386$ and 
$\gamma = 0.683$, for
the data shown in Fig. 1. $f^*$ fluctuates about a constant value 
over the optimum fitting interval, which in this case
turns out to be $[2.8,10.3]$.  (The derivative $\tau$ evaluated as in
Fig. 1 varies between
1.18 and 1.31 on this interval.)  For comparison we show the result of
a pure power-law fit using the estimate $\tau_s = 1.25$
\cite{biham}.  The latter yields a strongly curved function $f^* (x)$,
showing the inadequacy of a simple power law.

The best-fit parameters for sizes of nondissipative avalanches in
two dimensions are listed in Table I, with the final row indicating
the result of an extrapolation to infinite size (the data fall
close to a straight line when plotted versus $L^{-1/2}$).  The parameters
vary considerably with $L$, but in a systematic manner.  Our
estimate $\tau_{s,n} = 1.30(1)$ (here the subscript denotes size,
nondissipative; figures in parentheses denote uncertainties), 
is consistent with previous
estimates of 1.28(2) \cite{manna},
1.25(2) \cite{biham}, 1.27(1) \cite{chessa} and
1.28(1) \cite{pastor}.
(Note that these studies include both dissipative and
nondissipative avalanches in the analysis, which leads to a smaller exponent
estimate, since $\tau$ is smaller for dissipative avalanches than for
nondissipative ones.)  Another important
conclusion from the data in Table I is that the logarithmic
correction {\it does not dissappear} as $L \to \infty$.
The asymptotic avalanche distribution, while scale invariant, does not
follow a simple power law.

The duration distribution for nondissipative avalanches in 2-d,
and both size and duration distributions in the one-dimensional case,
follow the pattern described above.  In each instance 
the best-fit parameters $\tau$ and $\gamma$ decrease 
systematically with $L$, leading to
the exponent estimates in Table II.  Also listed are the exponents
$D$ governing the mean size and duration, defined via
$\overline{s}_n \sim L^{D_{s,n}}$ (and similarly for the mean
avalanche duration $\overline{t}_n \sim L^{D_{d,n}}$).  

For {\it dissipative} avalanches some intriguing differences appear.
In two dimensions, the size distributions can be fit to high accuracy
using $\gamma = 0.5$, and the duration distributions using $\gamma =1$.
Thus the logarithmic correction is generally weaker than
in the nondissipative case, and shows no significant size dependence.
This is reminiscent of the observation of clean power laws
for dissipative avalanches (but not for nondissipative ones) in the
BTW model \cite{drossel}.  In one dimension, {\it no} logarithmic 
correction is required to fit the dissipative avalanche distributions.
(In all other cases, $\gamma$ approaches a nonzero limiting value
as $L \to \infty$).  Finally, the best-fit values for $\tau$ show much less
size dependence than in the nondissipative case.  For avalanche sizes
in 2-d, for example, we find $\tau = 1.004$, 0.988, 0.985, 0.978, and 0.975,
for $L=160$, ...,2560, respectively.  In one dimension the exponent estimates
for dissipative avalanches show no systematic size dependence. 

Exponent values for dissipative avalanches are listed in Table III.
Since $\tau < 1$ for dissipative avalanches (as is also the case for
the BTW model \cite{drossel}), the distributions $P_s$ and $P_d$ will
not be normalizable if the scaling function $f$ 
in Eq. (\ref{logcr}) attains an $L$-independent
limiting value as $L \to \infty$.  In fact, we find that 
$f_0 \sim L^{-0.8}$ in one dimension, and $\sim L^{-0.2}$ in 2-d.

We have also performed a fitting analysis allowing the value of
$s_0$ (mentioned in the discussion following Eq. (\ref{logcr})) to vary.
While inclusion of an additional parameter leads to marginally 
improved fits, the best-fit values of $s_0$ do not differ greatly
from unity, and follow no systematic trend with system size.
(In the nondissipative case we tried fixing $\gamma =2$
or 3, and allowing $s_0$ to vary; there is no significant improvement
in the fit.)
Thus we find no advantage to including $s_0$ as a further adjustable parameter.

An important consequence of our results is that the Manna model
belongs to a different universality class than the BTW sandpile.
This follows by comparing the two-dimensional
Manna values, $\tau_{s,d} = 0.98(2)$ and $D_{s,d} =2.74(6)$,
with the known values of $\tau_{s,d} =7/9$ abd 
$D_{s,d} = 2$ for BTW \cite{drossel}.
Our findings strengthen
the conclusion reached by Biham et al. \cite{biham}, on the basis 
of SOC sandpiles, and by Vespignani et al. \cite{fes2}, who studied
``fixed energy" sandpiles.

In the 2-d case, the exponents $\tau_{s,d} $ and $\tau_{d,d}$ are so
similar, and so near unity, that one might conjecture that
they are both equal to 1.  It is therefore useful to determine
the relation between sizes and durations of dissipative avalanches,
which is expected to follow a power law, $t \sim s^{x_d}$.  A study
using $L=2560$ yields $x_d = 0.57(1)$.
Recalling the scaling relation
$x = (1-\tau_s)/(1-\tau_d)$ \cite{manna}, we see that 
$\tau_{d,d} < \tau_{s,d}$ in this case.
(The values quoted in Table III yield $x_d = 0.6(6)$; the uncertainty is
too great to permit a meaningful comparison.)
In the one dimensional case, $\tau_{d,d}$ and $\tau_{s,d}$ 
are sufficiently different from unity that a quantitative
comparison is possible.  Simulations ($L = 10^4$) yield
$x_d = 0.681(5)$, while $(1-\tau_s)/(1-\tau_d) = 0.682(6)$.
For nondissipative avalanches the comparisons are, in 1-d,
$x_n = 0.692(5)$ (simulation) and 0.47(27) (scaling); in 2-d,
$x_n = 0.593(5)$ (simulation), 0.55(6) (scaling).

In Figs. 3 and 4 we show size distributions for dissipative 
avalanches in 1d and 2d, respectively.
We find that finite-size scaling holds for dissipative avalanches in the Manna model,
unlike the BTW sandpile \cite{drossel}.  
In each case, we achieve a clean
data collapse by scaling the size $s$ to its mean, which is proportional to
$L^{D_{s,d}}$.  In one dimension, the probability must be multiplied
by $L^{D_{s,d}}$ to ensure collapse, so that the avalanche distribution
has the scaling form $P_s(s) = L^{-D} {\cal P}(L^{-D}s)$, with ${\cal P}$
a normalized scaling function.  In two dimensions $P$ must be 
multiplied by $L^{2.88}$ to obtain a collapse.

In summary, we find that for Manna's stochastic sandpile, simple
power-law avalanche distributions are the exception rather than the rule,
and are observed only for nondissipative avalanches 
in the one-dimensional system. (Since our conclusions are based entirely on
simulation data, we cannot rule out other, slowly decaying correction
to scaling forms.)    
Our results for $\tau_{s,d}$ and $D_{s,d}$ clearly place 
the Manna and BTW models in distinct universality classes.  

Our findings raise several questions regarding avalanche distributions.  
First, What is the physical origin and theoretical basis for 
the logarithmic correction?
Second, Do such corrections appear in other models exhibiting SOC?
Finally, Which are the essential features of the Manna and BTW models
leading to their rather different scaling properties?
We hope to investigate these issues in future work.
\vspace{2em}

\noindent{\bf Acknowledgements}
\vspace{1em}

We thank Deepak Dhar, Alessandro Vespignani and Kim Christensen
for helpful comments.
This work was supported by CNPq and CAPES, Brazil.

\newpage

\begin{table}
\begin{center}
\begin{tabular}{|r|c|c|}
\hline
 $L$   &   $\tau$   & $\gamma$   \\
\hline
 160   &  1.570  &  1.252 \\
 320   &  1.486  &  1.027 \\
 640   &  1.425  &  0.821 \\
1280   &  1.389  &  0.696 \\
2560   &  1.364  &  0.592 \\
\hline  
ext.   &1.302(10)&  0.356(10) \\
\hline
\end{tabular}
\end{center}
\label{tab1}
\noindent{Table I. 
Best-fit parameters for the distribution of sizes of nondissipative 
avalanches in two dimensions. The
final line gives estimated values for $L \to \infty$.}  
\end{table}

\begin{table}
\begin{center}
\begin{tabular}{|r|c|c|c|}
\hline
   case  &      $\tau$    & $\gamma$  & $D$    \\
   \hline
   s, 1-d    &  1.09(1)   &   0.70(9) & 1.92(1) \\
   d, 1-d    &  1.19(10)  &   1.54(5) & 1.23(1) \\
   s, 2-d    &  1.30(1)   &   0.36(1) & 1.94(2) \\
   d, 2-d    &  1.55(4)   &   0.85(6) & 0.72(1)  \\
\hline
\end{tabular}
\end{center}
\label{tab2}
\noindent{Table II. 
Best estimates for exponents associated with distributions of 
sizes (s) and durations (d) of  nondissipative avalanche distributions  
in one and two dimensions.}
\end{table}

\begin{table}
\begin{center}
\begin{tabular}{|r|c|c|c|}
\hline
   case  &   $\tau$    & $\gamma$ &  $D$    \\
   \hline
   s, 1-d  & 0.638(2)  &     0    & 2.20(1)  \\
   d, 1-d  & 0.469(2)  &     0    & 1.47(1)  \\
   s, 2-d  & 0.98(2)   &    1/2   & 2.74(6)  \\
   d, 2-d  & 0.965(5)  &     1    & 1.42(1)  \\
\hline
\end{tabular}
\end{center}
\label{tab3}
\noindent{Table III. 
Best estimates for exponent $\tau$ associated with distributions 
of sizes (s) and durations (d) of  dissipative avalanche distributions  
in one and two dimensions, with $\gamma$ fixed at the
indicated value.  Note the absence of a logarithmic correction in 1-d.}
\end{table}

\newpage
\noindent FIGURE CAPTIONS
\vspace{1em}

\noindent FIG. 1. Main graph: avalanche size distribution for
the two dimensional model, $L=1280$; data points: simulation;
smooth curve: cubic fit to the data on the
interval $4 < \ln s < 13$.  Inset: slope $-\tau$
of the cubic fit, versus $\ln s$.
\vspace{1em} 

\noindent FIG. 2. Plot of $f^* = s^\tau P_s(s)/(\ln s)^\gamma$
versus $\ln s$ for the data shown in Fig. 1. Lower curve:
best-fit for $3 < \ln s < 10$ using $\tau = 1.385$ and $\gamma = 0.672$;
upper curve: pure power-law fit using $\tau = 1.25$.
\vspace{1em} 

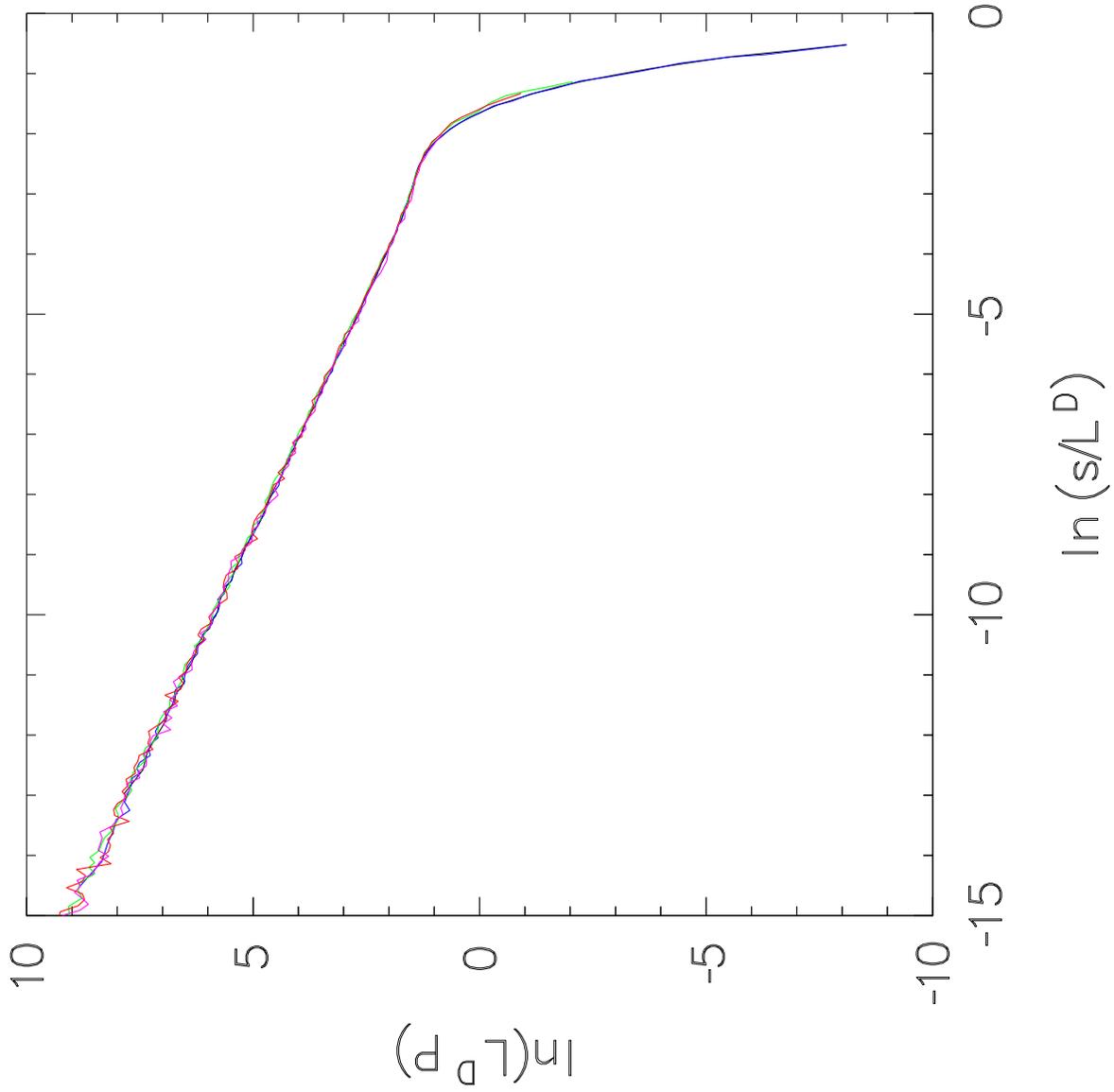
\noindent FIG. 3. Finite size scaling plot of
dissipative avalanche size distributions in
one dimension, $L=500$,...,$10^4$.
\vspace{1em} 

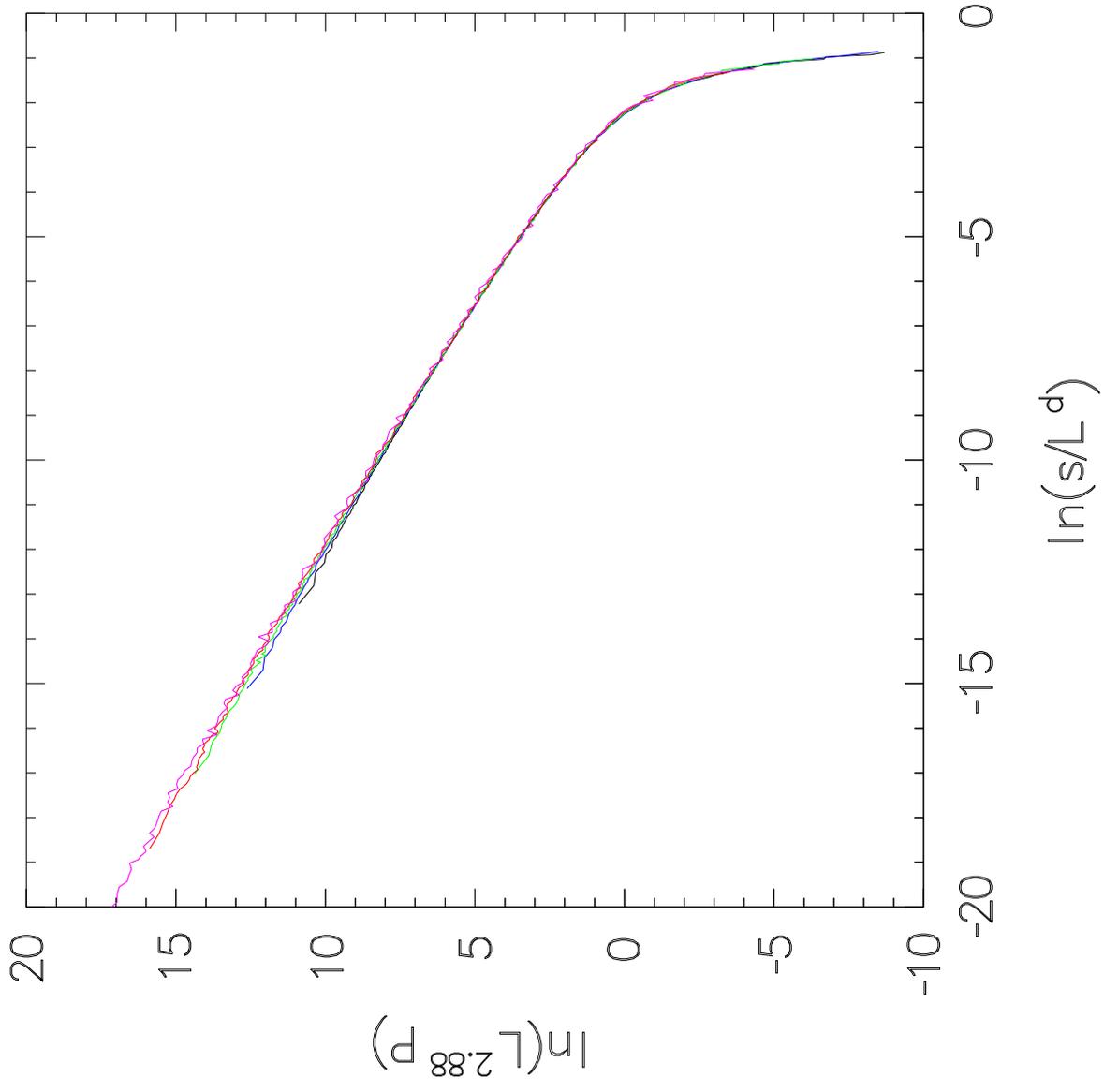
\noindent FIG. 4. Finite size scaling plot of
dissipative avalanche size distributions in 
two dimensions, $L=160$,...,2560.
%\vspace{1em}

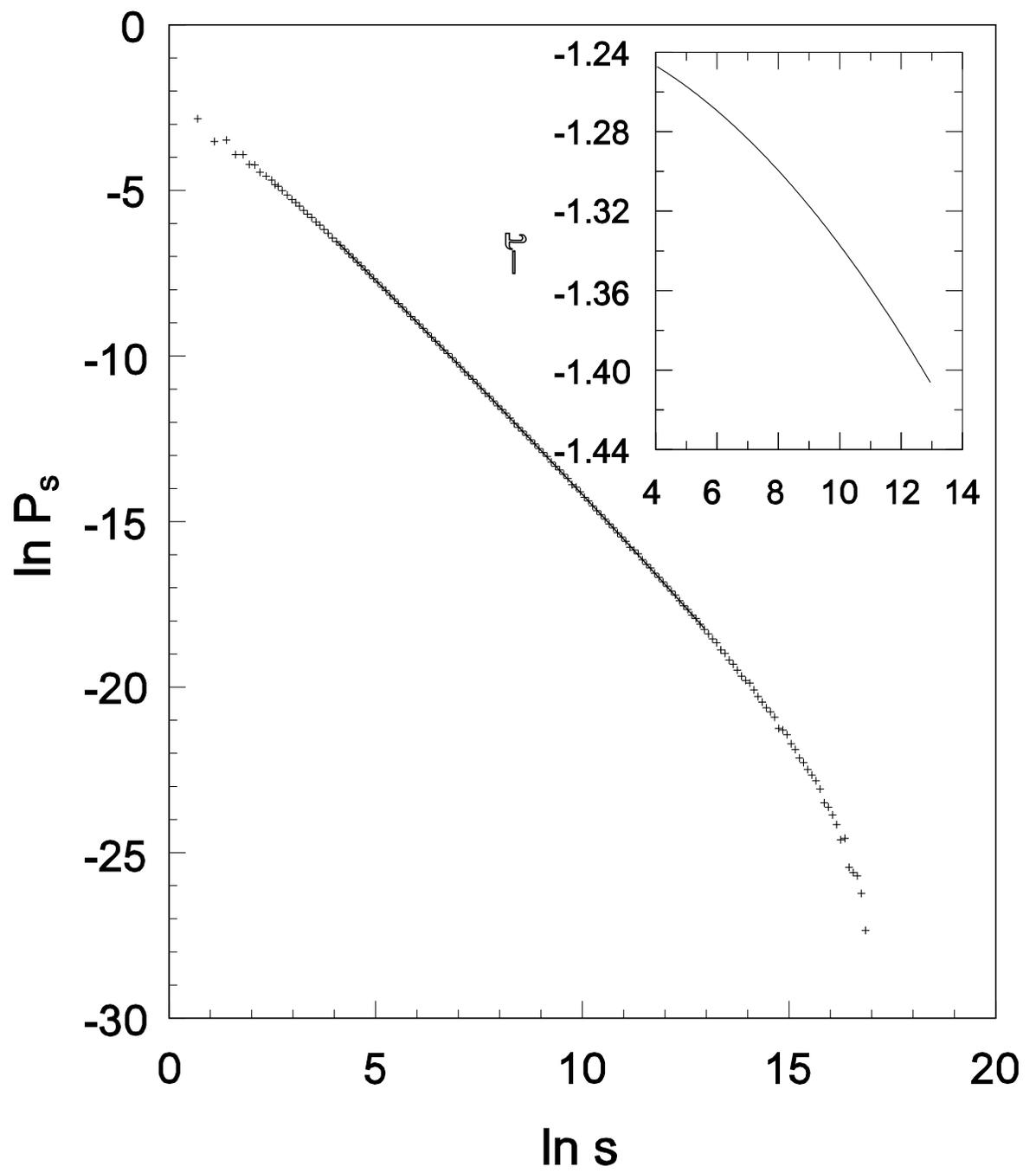
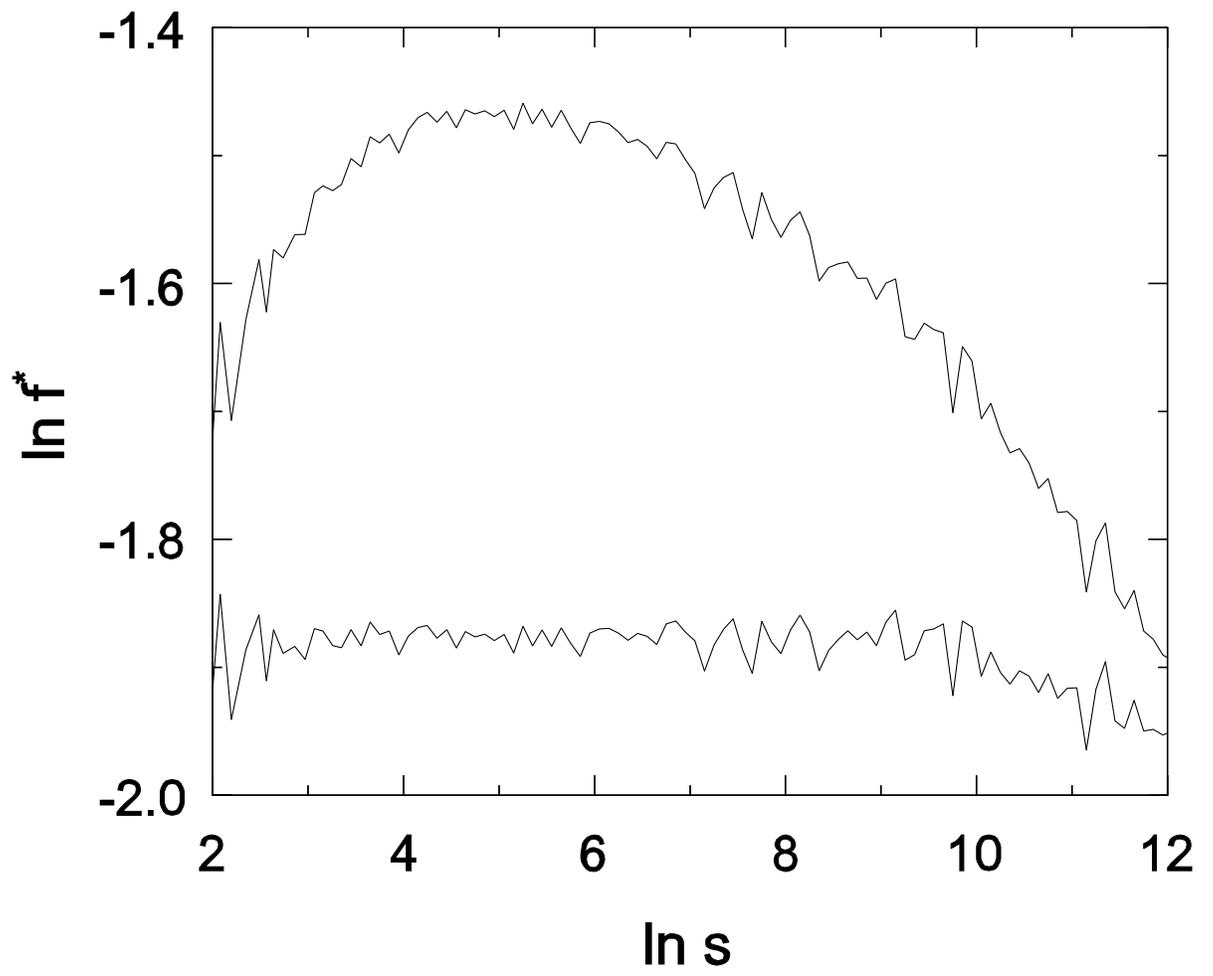
\end{document}